# FIRST-PASSAGE TIME: A CONCEPTION LEADING TO SUPERSTATISTICS. II. CONTINUOUS DISTRIBUTIONS AND THEIR APPLICATIONS

## V. V. Ryazanov*

*Institute for Nuclear Research, pr. Nauki, 47, 03068, Kiev, Ukraine*

**Abstract**

A continuous approximation for the results of [1] is obtained. In this approximation the energy distribution is represented in the form of the product of the Gibbs factor and superstatistics factor. The mutual weights of the factors are defined by the control parameter of the problem. Various approximations for the superstatistics factor are written. The resulting distribution is compared to a number of known results (multiplicative noise, Van der Pol generator etc). It is applied to the description of self-organized criticality, statistics of cosmic rays etc as well.

----------------------

*Corresponding author. E-mail address: vryazan@kinr.kiev.ua*

# 1. Continuos approximation for the superstatistics factor

In [1] the expression is obtained ((21)-(22) [1]):

$$p(E) = \frac{e^{-\beta E}}{Z(\beta,\gamma)} \omega(E) \sum_{k=1}^{n} R_k e^{-r_k E}, \qquad (1)$$

$r_k = \alpha_k y_k \Delta/u = \alpha_k[\beta_{0k}P_{0k}v_{0k} - \beta_{\gamma k}P_{\gamma k}v_{\gamma k}]/u$; $<r_k> = r_0 = \alpha_0 \Delta/u$; $\Delta = <\beta Pv>_0 - <\beta Pv>_\gamma$, $<y_k> = 1$. Having denoted $r_1 = = \alpha`y`\Delta/u$, $y_k \to y`$, $\alpha_k \to \alpha`(x)$, $R_k \to R`(x)$, $f(x) = dR`(x)/dx$; $v_k \to v`(x)$, $\beta_l = 1/k_B T`(x)$, $T_k \to T`(x)$, and proceeding to continuous sampling space, from discrete probabilities to continuous probability density function, and replaced summation by integration and defining density of probability $f(y_1 = r_1/r_0) = (dR`(x)/dx)/(d(r_1/r_0)/dx)$, we obtain from (1):

$$p(E) = exp\{-\beta E\}\omega(E)Z^{-1} \int_0^\infty f(y_1)exp\{-y_1 r_0 E\}d(y_1); \qquad \int_0^\infty f(y)dy = 1. \qquad (2)$$

Here we assumed that $f(r_1/r_0)$ is the probability density for a random $y(x) = r(x)/r_0$ to take on a specific value $y_1 = r_1/r_0$ under the condition that a system is located at the point $x$ of the potentially possible states considered as a continuous random value (although the initial set of potential values is a discrete set). Like in [1], we wrote $[\beta_{0k}P_{0k}v_{0k} - \beta_{\gamma k}P_{\gamma k}v_{\gamma k}]/u = r_k$, $\alpha = 1$, and represented the exponent in (2) as $-y_1 r_0 E$, $y_1 = (r_1/r_0)$, $r_0 = \Delta/u$, taking $r_0 E$ the argument of the characteristic function. Then the integrand in the rhs of (2) is the Laplas transform for the distribution $f`(y_1)$ with argument $r_0 E$.

The averaging over $r_1$ can be viewed as well from the isothermic-isobaric $f_V(p,q) = Q^{-1}(\beta,P,N)exp\{-\beta(H(p,q)+PV)\}$; $Q = \int exp\{-\beta(H(p,q)+PV)\}dzdV$; $H = E$ is the energy [2], where the variable $V$ (volume) fluctuates. In open system with the potential of a kind *Fig.2* [1] an extra factor $\Delta = <\beta Pv>_0 - <\beta Pv>_\gamma$ appears, accounting the fluctuations of the inverse temperature $\beta$ and values of $P$, $v$. It is supposed in (1)-(2) that the energy $E$ and volume $V$ are random extensive quantities, but their ratio (energy density) $u = E/V$ is nonrandom intensive variable. Like in (8) of [1] we assume $\beta = 1/k_B T = \beta_{Gibbs}$; value $\beta = 1/k_B T$ corresponds to average inverse temperature of full system; $k_B T$ characterizes the conveniently averaged energy for full system. It is assumed that $\beta$ does not generally coincide with $<\beta> = \beta^0 = \int \beta_l f(\beta_l r_l)d(\beta_l r_l)$. Similarly we assume that the pressure $P$ for the total system does not coincide with the averaged pressure $P_0 = <P> = \int P_l f(\beta_l r_l)d(\beta_l r_l)$ with averaging over all subsystems, and $r_f = [\beta_0 P_0 - \beta_\gamma P_\gamma]/u$ for the total system, where $P$ is pressure for the total system, need not coinciding with average $r_0 = <r> = \int rf(r)dr$, where $u = E/V$ ($R_k$ from (16), (18), (21), (22) of [1]), the examples of the values $\gamma$ from (5)-(7), (12)-(17) [1] are given at the end of the Section 3 of [1]; $\lambda_k$ from (20) [1] is intensity of energy flow in the system (subsystem), equal in dynamical equilibrium to an output intensity [3]. If in (2) we consider the system without fluctuations of value $r$ and set $f(r_1/r_0) = \delta(r_1/r_0 - 1)$, then $p(E) \sim exp\{-(\beta+r_0)E\}$. In this case assume that $p(E) \sim exp\{-\beta^0 E\}$, and thus we get $\beta^0 = \beta + r_0$, $\beta = \beta^0 - r_0$. If $\beta^0$ weak depended from $\gamma$, then $r_0 \approx \beta^0 p_0/u$; $p_0 = <Pv>_0 - <Pv>_\gamma$; $\beta = \beta^0(1-p_0/u)$; $\beta^0 = \beta/(1-p_0/u)$.

For a case of one class of ergodic states, when $lnQ = \beta PV$, $p(E) \sim exp\{-(\beta+r_f)E\}$, $p(E) \sim exp\{-\beta^0 E\}$, one gets $r_f = r_0$. Replacing $\beta_l$ by $\beta_l = 1/k_B T` = 1/k_B T_l(r,t)$, the fluctuating value of temperature, we get:

$$p(E)=\exp\{-\beta E\}\omega(E)Z^{-1}\int_0^\infty f(y_1)\exp\{-y_1 r_0 E\}d(y_1) = \qquad (3)$$

$$=f_B f_A/Z=p_B p_A(Z_B Z_A/Z); \quad Z=\int f_B f_A dE; \quad Z_A=\int f_A dE; \quad Z_B=\int f_B dE; \quad p_B=f_B/Z_B; \quad p_A=f_A/Z_A;$$

$$f_B(E)=\exp\{-\beta E\}\omega(E); \qquad f_A(E)=\int_0^\infty f(y_1)\exp\{-y_1 r_0 E\}d(y_1), \qquad y_1=r_1/r_0,$$

where $f_B$ is the Boltzmann factor, $f_A$ and $p_A$ corresponds to the type-$A$ superstatistics from [4]. For $f(y_1)=\delta(y_1-1)$ we obtain from (3) the ordinary Boltzmann factor. Equality $r_1=r_0$ is possible at $\beta_1\approx\beta_0$, $V_k P_k\approx VP$. As in [5], we can write $p(E)=\exp\{-\beta^0(1-p_0/u)E\}\omega(E)Z^{-1}\exp\{-r_0 E\}[1+\sigma_y^2(r_0)^2 E^2/2+O(\sigma^3 E^3)]=\exp\{-\beta^0 E\}\omega(E)Z^{-1}[1+(q`-1)(r_0)^2 E^2+g(q`)(r_0)^3 E^3+...]$, where $(q`-1)=\sigma_y^2$, $q`=<y^2>$; $(q`-1)^{1/2}=\sigma_y$; $\sigma_y^2=<y^2>-1$; value $\beta$ from [5] replaced by $r_1$, the fluctuating intensive parameter is equal to $r_1$ instead of $\beta$ as in [5]; $r_0=<r_1>=\int r_1 f(r_1)d(r_1)$, function $g(q`)$ depends on the superstatistics chosen [5]. For $f=f_G$ from (17) [1], (2), $q`=q=q_{Ts}$ [6,7]. For others $f$: $q`=q_{BC}$ [5]. For any distribution $f(r_1)$ with average $r_0=<r_1>$ and variance $\sigma^2$ we can write $p(E)\sim\exp\{-\beta^0(1-p_0/u)E\}<\exp\{-y_1 r_0 E\}>=\exp\{-\beta^0 E\}<\exp\{-(y_1-1)r_0 E\}>=\exp\{-\beta^0 E\}[1+\sigma_y^2(r_0)^2 E^2/2+\sum_{r=2}^\infty (-1)^r$

$<(y_1-1)^r>(r_0)^r E^r]$ similarly [5]. Thus, low energy asymptotics of the obtained distribution (3) coincides with the asymptotics of superstatistics [8]. But the high energy asymptotics differs from the asymptotics obtained in [8] (*Fig. 3*).

If one performs the replacement of variables $\beta`=\beta+r_1$ and assumes that the situation $\beta=0$ is possible, when the bottom limit of integration in (3) after replacement of a variable is $0$, instead of (3) we shall obtain that $p(E)=\omega(E)Z^{-1}\int_0^\infty f_1(\beta`)\exp\{-\beta` E\}d\beta`$, $f_1(\beta`)=f(r_1)=f[\beta`-\beta^0(1-p_0/u)]$, that coincides with superstatistics [5]. But the correlation (3) describes a more general situations and superstatistics forms here a special case (3). For ideal gas $P/u\approx 0,687$.

As in [1] by $\alpha=1$

$$p(E) = \frac{e^{-\beta E}}{Z(\beta,\gamma)}\omega(E)\int_0^\infty f(y_1)[1+\frac{\gamma_1}{\lambda_1}\exp\{(\beta PV)_{01}\}]^{-1}dy_1.$$

This distribution tends to Gibbs one by $\gamma\to 0$. By big values $\gamma$, $p(E)\to 0$. By $\gamma\approx\lambda$,

$$p(E)\sim\frac{e^{-\beta E}}{Z(\beta,\gamma)}\omega(E)\int_0^\infty f(y_1)\exp\{-(\beta Pv)_{01} E/u\}dy_1.$$

It is possible to mention some examples of physical realization of distribution (3). So, for a case of superdiffusion in [9] at the absence of external force stationary distribution is in Tsallis form, and for a case of presence of constant external force and multiplicative noise in exponential form of distribution is manifested. As in (3), in one phenomenon for one system two kinds of distribution are combined (see also Sections 3, 4).

Besides the stated scheme there are also other opportunities of realizing the distributions such as (3) and generalizations of the theory of superstatistics. Except for distributions (17) - (18) [1] for density of probability of lifetime in the structural factor $\omega(E,\Gamma)$ (16) [1] can consider other distributions as well. Since the phase space of complex systems gets intricate fractal structure [10] it is natural to build a distribution on a fractal substrate. As distribution $f$ in (3) it is possible to choose a fractal distributions as in [11] or their generalizations on a case of multifractals, and also other distributions used in [5]. Other various combinations of different distributions in the different purposes are possible as well. The obtained distribution of a kind (3) gives the wide opportunities of the description of various physical situations.

In [12] it is shown that the temperature fluctuations can be described by the gamma-distribution. As in [5] for distribution function $f$ in (3) as gamma-distribution such as (17) [1] (but with variables $\alpha_k=c$, $b_k=b$; $f(y)=\Gamma^{-1}(c)b^{-c}(y)^{c-1}\exp\{-y/b\}$) expression (3) reduces to the product of Gibbs distribution and Tsallis distribution of a kind

$$p(E)=\exp\{-\beta E\}\omega(E)Z^{-1}(1+br_0E)^{-c}=$$
$$=\exp\{-\beta E\}\omega(E)Z^{-1}[1+(q-1)r_0E]^{-1/(q-1)}; \quad c=1/(q-1); \quad b=q-1. \qquad (4)$$

At $q\to 1$ one gets from (4) the expression $\exp\{-\beta^0 E\}$, which at $r_0=0$ yields Gibbs distribution $\exp\{-\beta E\}$. In (4) we can separate the Tsallis temperature, $\beta_q=\beta_{Ts}=r_0$. Thus we do have three different "temperatures": Gibbs one (8) from [1] $\beta=\beta_{Gibbs}=1/k_BT=\beta^0(1-p_0/u)$; effective temperature $\beta^0=\beta_{eff}=\beta/(1-p_0/u)$; and Tsallis temperature $\beta_q=\beta_{Ts}=r_0=\beta(p_0/u)/(1-p_0/u)$. Following relations between them and $r_0$, $r_f$ values hold:

$$\beta^0=\beta+\beta_q; \qquad r_f=r_0=\beta^0-\beta=\beta_q. \qquad (5)$$

Similar conjectures can be obtained by means of the isothermic-isobaric ensemble as well [2]. For other distributions different from (17) [1] and (4) we can get different expressions for the corresponding temperatures $\beta_q$. In such a manner besides the choice of the distribution parameters (17) of [1] made in [1] and (4), in [4,13] one averages the value $(\beta/2\pi)^{1/2}\exp\{-\beta u^2/2\}$ using the $\chi^2$–distribution; in this case the factor $(\beta/2\pi)^{1/2}$ results in the change of the parameter $b$ in (4). For the Tsallis distribution one gets $\beta_{Ts}=\beta_q=\tilde{\beta}=2r_0/[2-(q-1)]$. Correspondingly one should set $bc=\tilde{\beta}/r_0=2/[2-(q-1)]$ for this case.

It is obvious that distribution (3) represents the product of Boltzmann factor with a multiplier superstatistics from which Tsallis distribution as a special case can be obtained (having chosen $f$ in (3) the gamma-distribution of a kind (17) [1]). In *Fig. 1* it is shown the comparison of distribution (4) $f(v)=\exp\{-\beta v^2/2\}[1+(q-1)r_0v^2/2]^{-1/(q-1)}/Z$ ($E=v^2/2$; $\beta=0,1$; $r_0=0,1$; $q=1,1$) with distribution (23) from [1] $p(v)=\exp\{-\beta v^2/2\}[1+(q-1)(1-\exp\{-r_0v^2/2\})]^{-1/(q-1)}/Z_1$, with Boltzmann-Gibbs distribution $n(v)=\exp\{-\beta v^2/2\}/Z_B$ and with Tsallis distribution $t(v)=[1+(q-1)\beta(p_0/u)v^2/2(1-p_0/u)]^{-1/(q-1)}/Z_T$. It is seen that the distributions $f(v)$ from (4) and $p(v)$ from (23) of [1] coincide and lie between the Boltzmann-Gibbs $n(u)$ and Tsallis distribution $t(v)$, although their "tails" (*1,339×10⁻¹⁰* for $p(v)$ and *4,245×10⁻¹¹* for $f(v)$ at $u=-20$) have values below the corresponding tail values of $n(v)$ (*6,917×10⁻¹⁰*) and $t(v)$ (*5,916×10⁻³*).

The behaviour of distribution (3), (4) essentially depends on values $r_0$. At $r_0\to 0$ the Gibbs distribution is obtained. As it is marked in the beginning of Section 3 [1], it corresponds to transition to Gibbs distributions.

## 2. Various models of superstatistics

Likewise for the discrete case [1] one can specify the expressions obtained. As $f$ in (2)-(3) one can choose various continuous expressions, for example, a functional gamma-distribution with density of the form $g(x)=\lambda^\alpha \beta^\alpha [ln(1+x)]^{\alpha-1} exp\{-\lambda\beta ln(1+x)\}/\Gamma(\alpha)(1+x)$ and parameters $\alpha$, $\beta$, $\lambda$.

As in [1] one can use the superposition of the characteristic functions, i.e. one characteristic function serves as an argument for another: $\varphi(f(t))$. Multilevel hierarchy $\varphi(f_1(f_2(...)))$ can be considered as well. Using in $\varphi(f(t))$ gamma-distributions and relations (4) for both $\varphi$ and $f$ yields the distributions $p(E)$ of the form:

$$p(E)) \sim exp\{-\beta E\}[1+(q-1)ln\{1+(q_1-1)r_0E\}/(q_1-1)]^{-1/(q-1)}. \qquad (6)$$

At $q_1 \to 1$ we get the expression (4). If $q \to 1$ the same expression (4) is recovered, where $q$ is replaced by $q_1$.

For $\varphi(f_1(f_2(t)))$ and using gamma-distributions with (4) for $\varphi$ and $f_1, f_2$ one gets:

$$p(E)) \sim exp\{-\beta E\}[1+(q-1)ln\{1+(q_1-1)ln[1+(q_2-1)r_0E]/(q_2-1)\}/(q_1-1)]^{-1/(q-1)}. \qquad (7)$$

For logarithmic-normal distribution for the function $f$ in (2)-(3) with density $g(x)=exp\{-(lnx-\mu)^2/2b^2\}/xb(2\pi)^{1/2}$, $x \geq 0$, the characteristic function acquires the form:

$$\varphi(t) = 1 + \sum_{k=1}^{\infty} (it)^k exp\{\mu k+k^2b^2/2\}/k!,$$

and

$$p(E) \sim exp\{-\beta E\}[1+\sum_{k=1}^{\infty}(-r_0E)^k exp\{\mu k+k^2b^2/2\}/k!]. \qquad (8)$$

This can be compared to Tsallis distribution (4) obtained out of gamma-distribution, when

$$p(E) \sim exp\{-\beta E\}[1+\sum_{k=1}^{\infty}(-r_0E)^k[(k-1)q-(k-2)]/k!].$$

Using as $f$ in (2)-(3) the beta-distribution of the second kind with density $g(x)=(a/b)^\alpha x^{\alpha-1}/B(\alpha, c)(1+ax/b)^{\alpha+c}$, $x \geq 0$ ( $\alpha$, $c$, $a$, $b$ – parameters and $B(\alpha, c)$ is beta-function) and with characteristic function $\varphi(t)=\Gamma(\alpha+c)\Psi(\alpha,1-c;-itb/a)/\Gamma(c)$, where $\Psi$ is the degenerated hypergeometric function and $\Gamma$ is gamma-function, leads to the energy distribution of the form

$$p(E) \sim exp\{-\beta E\}\Gamma(\alpha+c)\Psi(\alpha, 1-c; br_0E/a)/\Gamma(c), \qquad (9)$$

This latter at big $E$ yields the asymptotics $p(E) \sim exp\{-\beta E\}(r_0E)^{-\alpha}$. The degenerate hypergeometric function $\Psi$ is contained also in the Fisher distribution, noncentral $t$-distribution $II$, $t^2$- and noncentral $t^2$-distributions, $F$-distribution with characteristic function $\varphi(t)=\Gamma((v+w)/\alpha)\Psi(v/2, 1-w/2; -itw/v)/\Gamma(w/\alpha)$ and corresponding distribution function

$$p(E) \sim exp\{-\beta E\}\Gamma((v+w)/\alpha)\Psi(v/2, 1-w/2; wr_0E/v)/\Gamma(w/\alpha) \qquad (10)$$

(we note that in [5] the $F$-distribution is considered, but the explicit form of the characteristic function is unspecified, hence non specified is the closed expression for $p(E)$); noncentral $F$- distribution, Meykhem-distribution etc. The combinations of the degenerate hypergeometric functions $\Psi$ contain noncentered Beta-distribution of the second kind and other types of distributions. Equally many other characteristic functions contain special functions. Thus characteristic function of the standard Beta-distribution of the second kind contains the Whitteker function; the characteristic functions of the logistic, Nakagami-, and $\chi$-distribution contain the functions of the parabolic cylinder.

For the noncentered gamma-distribution with density $g(x)=e^{-\delta}g_1(\alpha, \lambda, x)_0F_1(\alpha, \delta x/\lambda)$ ($\alpha>0$, $\lambda>0$, $\delta\geq0$ are parameters, $g_1(\alpha, \lambda, x)$ the density of gamma (17) of [1] provided that $\lambda=1/b_k$, $\alpha=\alpha_k$, $_pF_q$ is the generalized hypergeometric series) the characteristic function is $\varphi(t)=(1-it\lambda)^{-\alpha}exp\{it\delta\lambda/(1-it\lambda)\}$. It generalizes the characteristic function of gamma-distribution and contains an extra parameter $\delta$. If $\alpha=1/(q-1)$, $\lambda=(q-1)/(1+\delta(q-1))$ then

$$p(E) \sim exp\{-\beta E\}[1+r_0E(q-1)/(1+\delta(q-1))]^{-1/(q-1)}exp\{-\delta(q-1)r_0E/[1+\delta(q-1)+(q-1)r_0E]\}.$$

The estimations show that at relatively small $\delta<0,9$ and for the values of parameters $r_0=0,1$; $\beta=0,1$; $q=1,1$ (as on *Fig. 1*) this distribution differs slightly from (4) (see *Fig. 1*). It is possible to express $\delta$ through a new nonextensive parameter $\tilde{r}_0 =r_0/[1+\delta(q-1)]$; $\delta(q-1)= (r_0/\tilde{r}_0)-1$, then

$$p(E) \sim exp\{-\beta E\}[1+(q-1)\tilde{r}_0 E]^{-1/(q-1)}exp\{-(r_0-\tilde{r}_0)E/[1+(q-1)\tilde{r}_0 E]\}. \tag{11}$$

It is possible to introduce the nonextensive effective inverse temperature also.

Note also the Pearson-*III* distribution with density (at $a>0$ $a\leq x<\infty$, $a$, $\alpha>-1$ are parameters) $g(x)=a^{(\alpha+1)}(1+x/a)^{\alpha}exp\{-a(1+x/a)\}/|a|\Gamma(\alpha+1)$, and characteristic function $\varphi(t)=(1+ita/2)^{-\alpha-1}exp\{ita\}$, for this case $p(E) \sim exp\{-\beta E\}[1-(a/2)r_0E]^{-\alpha-1}exp\{-ar_0E\}$. At $\alpha+1=1/(q-1)$, $a/2=-(q-1)$ this is the product of the Tsallis distribution and the factor $exp\{2r_0E(q-1)\}$. In order that $<r>=r_0$ holds, we take $a=2(q-1)/(2q-3)$; then

$$p(E) \sim exp\{-\beta E\}[1+(q-1)r_0E/(2q-3)]^{-1/(q-1)}exp\{2(q-1)r_0E/(2q-3)\}=$$

$$exp\{-\beta E[(1-2(q-1)r_0/(2q-3)(1-r_0)]\}[1+(q-1)r_0E/(2q-3)]^{-1/(q-1)}.$$

Characteristic functions for the Pearson distributions of the types *IV*, *V* and *VII* are expressed through the cylindric function with imaginary argument $K_v(z)$, the Pearson distributions of the *VI* type – through Whitteker's $W_{\lambda,\mu}(z)$. The Bessel distributions with density $g(x)=exp\{-(x-a)/\lambda\}\alpha J_\alpha((x-a)/\lambda)/(x-a)$, $x\geq a$, where distribution parameters are $|a|<\infty$, $\alpha>0$, $\lambda>0$, possesses a characteristic function $\varphi(t)=exp\{ita\}[1-it\lambda-((1-it\lambda)^2-1)^{1/2}]^\alpha$, and $p(E) \sim exp\{-\beta E-r_0Ea\}[1+r_0E\lambda-((1+r_0E\lambda)^2-1)^{1/2}]^\alpha = exp\{-\beta E-r_0Ea\}[1/2]^\alpha(1+r_0E\lambda)^{-\alpha}[1+1/4(1+r_0E\lambda)^2+1/8(1+r_0E\lambda)^4+...]^\alpha$. At $\lambda=q-1$, $\alpha=1/(q-1)$ we receive the product of the Tsallis distribution $[1+r_0E(q-1)]^{-1/(q-1)}$ and $exp\{-\beta E-r_0Ea\}[1/2]^{1/(q-1)}[1+1/4(1+r_0E\lambda)^2+ 1/8(1+r_0E\lambda)^4+...]^\alpha$. The average of the Bessel distribution diverges.

The compound exponential distribution is described by the density $g(x)=\int_0^\infty \lambda \exp\{-\lambda x\}f_1(\lambda)d\lambda$. In this case the parameter $\lambda$ of the exponential distribution is a random value with density $f_1(\lambda)$, and

$$p(E) \sim exp\{-\beta E\} \int_0^\infty [\lambda/(\lambda+r_0E)]f_1(\lambda)d\lambda.$$

If for $f_1(\lambda)$ one chooses the gamma-distribution $f_1(\lambda)=\lambda^{c-1}exp\{-\lambda/b\}/b^c\Gamma(c)$, then the Pareto distribution is yielded with characteristic $\varphi(-r_0E)=\Gamma(c+1)(r_0E)^c exp\{r_0E/b\}\Gamma(-c,r_0E/b)/\Gamma(c)b^c$. Choosing $c$ and $b$ like for Tsallis distribution (4), $b=q-1$, $c=1/(q-1)$, we get the expression

$$p(E) \sim exp\{-\beta E\}(r_0E)^{1/(q-1)} exp\{-r_0E/(q-1)\}\Gamma(-1/(q-1), r_0E/(q-1))/\Gamma(1/(q-1))(q-1)^{1/(q-1)}. \quad (12)$$

Since at big $z$, $\Gamma(a,z) \sim z^{a-1}exp\{-z\}[1+(a-1)/z+...]$, we have for large $E$: $p(E) \sim exp\{-\beta E\}\Gamma(q/(q-1))(q-1)(r_0E)^{-1}exp\{-r_0E/(q-1)\}[1-q/r_0E+...]/\Gamma(1/(q-1))$.

Hyper Erlang distribution $p(E) \sim exp\{-\beta E\} \sum_{i=1}^{N} \frac{\alpha_i \lambda_i^{n_i}}{(\lambda_i+r_0E)^{n_i}}$ at $\sum_{i=1}^{N}\alpha_i=1$, $\lambda_i=n_i=1/(q_i-1)$, describes the composition of Tsallis distributions:

$$p(E) \sim exp\{-\beta E\} \sum_{i=1}^{N} \frac{\alpha_i}{(1+(q_i-1)r_0E)^{1/(q_i-1)}} \,. \quad (13)$$

In conclusion we remark that besides the expressions (6)-(13) the options of modelling the explicit shapes of superstatistics are extremely vast, and their careful analysis, as well as their suitability to various physical problems are grateful subjects of the future challenges.

## 3. The role of the controlling parameter in the passage from superstatistics to Gibbs distributions and in the evolution of a system

The distributions (3), (4) are widely encountered in various domains of physics. We trace some examples with pointing out the role of the controlling parameters, which are included in the parameter $r_0$.

3 a) Distributions for the systems with multiplicative noise.

In [15] a family of the statistical models with multiplicative noise is considered. Microscopic dynamics, containing multiplicative noise, may be encountered in many dynamical processes, such as in stochastic resonance [16], noise induced phase transitions [17], granular packings [18], and others [19, 20] processes subject to both additive and multiplicative noises and described by the dimensionless stochastic differential equation of the form

$$du/dt = h(u) + g(u)\xi(t) + \eta(t), \quad (14)$$

where $u(t)$ is a stochastic variable, $h$, $g$ are arbitrary functions ($g(0)=0$), and $\xi(t)$, $\eta(t)$ are uncorrelated and Gaussian-distributed zero-mean white noises, hence satisfying

$$<\xi(t)\xi(t`)>=2M\delta(t-t`), \quad <\eta(t)\eta(t`)>=2A\delta(t-t`), \tag{15}$$

where $M$, $A>0$ are the noise amplitudes and stand for "multiplicative" and "additive", respectively.

The Fokker-Planck equation for the probability density $p(u,t)$, associated to Eq. (14) in the Stratonovich definition written in [15] as

$$\partial_t p=-\partial_u(h(u)p)+M\partial_u(g(u)\partial_u[g(u)p])+A\partial_{uu}p . \tag{16}$$

In [15] one considers various situations which as their partial cases lead to the distributions of the nonextensive statistical mechanics. In [15] it was noted that no functional exists where from by means of a variation procedure the distributions of the kind of (3)-(4) combining Gibbs and nonextensive statistical mechanics, could be derived. However in [1] the distribution (5) ([1]) was obtained merely from a variational procedure; further the distribution (3) was obtained therefrom under plausible assumptions. That is we got the distribution (3) in two steps: at first the variation procedure yielded the distribution (5) of [1], and then using assumptions as to the shape of the $\omega(E,\Gamma)$ function and integrating over $\Gamma$ we arrived at the distribution (3).

Under the zero flux boundary conditions (i.e. such that $j(-\infty)=j(\infty)=j(u)=0$, where $j(u)=J(u)p-\partial_u[D(u)p]$ is the current, $J(u)=h(u)+Mg(u)g`(u)$; $D(u)=A+M[g(u)]^2$) the stationary solution of (16) is written as

$$p(u)=C/\eta(u); \quad C=const; \quad \eta(u)=(Mg^2+A)^{1/2}exp\{-\int h(u)du/(Mg^2+A)\}. \tag{17}$$

If we assume the deterministic drift $h(u)$ and $g(u)$ in the form widely used in synergetics (for example, [21]):

$$h(u)=\alpha u-\delta u^3; \quad g(u)=u, \tag{18}$$

then from (17)-(18):

$$p(u)\sim exp\{-\delta u^2/2M\}/[1+Mu^2/A]^{[1-(\alpha+\delta A/M)/M]/2} . \tag{19}$$

Comparing (19) and (4), we get

$$E=u^2/2; \quad \delta/M=\beta=1/k_BT; \quad 1/(q-1)=[1-(\alpha+\delta A/M)/M]/2; \quad (q-1)r_0=2M/A;$$

$$r_0=(M-\alpha)/A-\beta; \quad \beta^0=\beta+r_0=(M-\alpha)/A=\beta+\beta_q; \quad \beta_q=r_0=(M-\alpha)/A-\beta=\beta_0-\beta.$$

From here it is clear that the relations (5) are satisfied, the sign of $q-1$, $r_0$ depends on the relation between $\alpha$, $\delta$, $A$, $M$. The value $\beta^0=\beta$ at $\alpha=M-\beta A$, $r_0=0$. The value $q-1$ has a singularity at $1=(\alpha+\delta A/M)/M$, changing its sign. At $\alpha=M$ the value $\beta^0$ turns to zero and the value $r_0=-\beta$. From hence it follows that the value of the controlling parameter $\alpha=M$ corresponds to the bifurcation point. Besides the (18)-(19) one can name a number of different examples of choosing functions $h(u)$ and $g(u)$ [15, 11]. As $f$ in (3) not only gamma-distributions (yielding Tsallis distribution (4)), but other possible classes of functions can be used.

3 b) Turbulence

In [22] a simple modification of distribution, obtained in [23] for acceleration $a$ of a tracer particle by the turbulent flow, is suggested for the description of the turbulence:

$$p(a)=C\exp\{-a^2/a_c^2\}/[1+V_0(q-1)a^2/2]^{1/(q-1)}; \quad V_0=4; \quad q=3/2, \quad (20)$$

where $C$ is normalization constant, $a_c$ is a free parameter, which in [22] is used for a fitting. In essence, this meant the redefinition $\beta/2 \to a_c^{-2}+\beta/2$. The numerical value $a_c=39,0$ is obtained in [22] by a direct fitting of the probability density function (20) to the experimental data. Direct comparison of (20) with (4) gives:

$$E=a^2/2; \quad \beta=2/a_c^2; \quad r_0=V_0; \quad \beta^0=V_0+2/a_c^2; \quad \beta_q=r_0=V_0=4. \quad (21)$$

Since in our case $\beta \approx 1,3 \times 10^{-3}$, then $\beta^0 \approx 4,0013 \approx V_0=4$.

The distribution and behaviour of this system is closer to the Tsallis distribution (consequently, to corresponding behaviour) than to Gibbs one.

3 c) Van der Pol generators

In [24] the Brownian motion in the autooscillating systems is considered on the example of the Van der Pol generator. This generator contains the linear electric oscillating contour included via the amplifier into the feedback unit. In [24] the situation is considered which is characteristics for the generation process in a number of cases (for example, for the theory of solid body lasers), where the dissipative nonlinearity enters symmetrically the equations for the charge and for the current (or coordinate and moment, speaking in terms of the mechanical analogy). The stationary distributions for the corresponding kinetic equation in [24] are written as

$$p_0(E)=C\exp\{-H_{eff}(E,\alpha_f)/k_BT\}; \quad H_{eff}=E-(\alpha_f/\delta)\ln(1+\delta E/\nu), \quad (22)$$

where $E$ is energy, $k_BT=D/\nu$ - effective temperature, $D$ – noise intensity, $\alpha=\alpha_f-\nu$, $\alpha_f$ – feedback coefficient, $\nu$ and $\delta$ are coefficients of linear and nonlinear friction; $\nu(E)=\nu+\delta E$, and the dependence of the diffusion coefficient $D(E)$ on energy $E$ has the form: $D(E)=\nu(E)k_BT/M=D(1+\delta E/\nu)$; $D=\nu k_BT$, here $M$ is the mass (inductivity in terms of the electromechanical analogy, we suppose $M=1$). We see that (22) coincides with (4) at

$$\beta=1/k_BT; \quad 1/(q-1)=-\alpha_f/\delta k_BT; \quad (q-1)r_0=\delta/\nu; \quad r_0=-\beta\alpha_f/\nu; \quad \beta^0=\beta(1-\alpha_f/\nu); \quad \beta_q=r_0=-\beta\alpha_f/\nu. \quad (23)$$

Let us consider the evolution of the system when the controlling parameter $\alpha_f$ changes. At $\alpha_f<0$, $r_0>0$, $r_f>0$, $\beta_q>0$, $\beta^0>0$, $q-1>0$. If $\alpha_f \to 0$ the value $q-1$ undergoes a peculiarity, and likewise $\beta_q$, $r_0$, changes its sign when the sign of $\alpha_f$ changes. If $\alpha_f-\nu<0$ the system tends to the state of rest with $E=0$, and $\beta^0>0$. At $\alpha_f=\nu$, $\beta_0$ becomes zero. The value $\alpha_f=\nu$ corresponds to the bifurcation point. At $\alpha_f-\nu>0$, $\beta^0<0$. The negative effective temperature corresponds to the selforganization, like in self-organized criticality (*SOC*) [25]. Stationary autooscillations emerge with the frequency $\omega_0$ and limit cycle energy $E=\alpha/\delta$ [24].

3 d) The velocity distribution in the presence of the heat source

The same form as (22) is acquired by the stationary solution of the kinetic equation in the presence of the heat source [24]. In [24] the following illustrative example is considered. Given the thermostate with the predefined temperature $T_0$ and provided with the heat source and sink. Linear coefficient $\nu$ (proportional to the energy $E$) and nonlinear one $\delta$ of the heat sink characterize the latter. The heat source is characterized by the feedback coefficient $a_f$. The "friction" is represented as

$$\nu(E)=\nu-a_f+\delta E; \quad E=mv^2/2; \quad \nu>0; \quad \delta>0.$$

The diffusion coefficient in the velocity space is nonlinear and given by the expression

$$D_{(v)}(E)= \nu(1+\delta E/\nu)k_BT/m.$$

Examining the dynamic equation for the energy it follows that the zero value of the parameter $a=a_f-\nu$ is the bifurcation point - $a<0$ the stable solution $E=0$, and at $a>0$ the nonzero solution $E=a/\delta$. Transiting through the bifurcation point a nonequilibrium phase transition occurs, and the structure of the distribution is changed. The controlling parameter is the feedback parameter $a_f$. The stationary distribution equals to

$$p(v)=C\exp\{-mv^2/2k_BT\}[1+(\delta/\nu) mv^2/2]^{a_f/k_BT\delta}. \tag{24}$$

Comparing (4) with (25), we get:

$$\beta=1/k_BT; \quad 1/(q-1)=-a_f/\delta k_BT; \quad (q-1)r_0=\delta/\nu; \quad r_0=-\beta a_f/\nu; \quad \beta^0=\beta(1-a_f/\nu); \quad \beta_q=r_0=-\beta a_f/\nu. \tag{25}$$

The expressions (25) coincide with (23) after replacement of $a_f$ to $\alpha_f$. This system exhibits all peculiarities characteristics for the Van der Pol generator if one takes the controlling parameter $a_f$ instead of $\alpha_f$.

3 e) Space diffusion of the Brownian particles in the external field

Besides the cases considered above the distributions (22), (24) suit for a number of other phenomena. So, for the spatial diffusion in the external field [24] $F=-gradU(r)$, if the elastic force becomes nonlinear and the potential is given $U(r)=(M\omega_0^2r^2/2)(-a+br^2/2)$, $a=a_f-1$, $b>0$. The coefficient $a_f$ characterizes the action of the effective field, for example, the Lorentz field in a dielectric [26] (if its value is big enough, at $a>0$, the elasticity coefficient becomes negative, and the state of the system turns out to bistable), and the coefficient $b$ is determined by the characteristics of the medium. The diffusion coefficient depends on the coordinates as $D_{(r)}(r)=D_{(r)}(1+br^2)$; $D_{(r)}=k_BT/M\nu$. The distribution obtained in [24] as the result of solving Einstein-Smoluchovsky equation is:

$$p(r)=C\exp\{-U_{eff}(r)/k_BT\}; \quad U_{eff}(r)=(M\omega_0^2/2)[r^2-(a_f/b)\ln(1+br^2)]. \tag{26}$$

Here an effective potential $U_{eff}$ is introduced which accounts for lowering the thermostat symmetry when increasing the coefficient $a_f$. This changing can take place when changing both density and temperature.

In this case $E=(M\omega_0^2/2)r^2$; $\beta=1/k_BT$; $1/(q-1)=-(M\omega_0^2/2)(a_f/bk_BT)$; $(q-1)r_0(M\omega_0^2/2)=b$; $r_0=-\beta a_f$; $\beta_q=r_0=-\beta a_f$.

The effective inverse temperature is $\beta^0=(1-a_f)/k_BT=\beta(1-a_f)$.

The resemblance to the cases 3c), 3d) is evident. Compared to 3d) the value $a_f/v$ is replaced by $a_f$. The value $a=0$, $a_f=1$ corresponds to the bifurcation point. The surrounding provides the nonlinearity of the force acting upon a Brownian particle and, as it is marked in [24], one could conventionally speak of the nonlinear thermostat. In [24] the difference in behaviour on the tails (at big $r$) is noted for the distribution (26) at $a_f\neq 0$ and $a_f=0$ (the latter coinciding with the Boltzmann distribution for the Brownian motion of harmonic oscillators).

3 f) Maltus-Ferhuelst process

The Maltus-Ferhuelst process for modelling the populations (e.g. bacteria's) with $N$ members is characterized by the death $v$ and birth $\alpha$ rates and extinction rate caused by interspecies struggle $\delta(N-1)$. In [24] the general stationary solution of the corresponding Fokker-Planck equation is given as

$$p_0(N,\alpha)=Cexp\{-H_{eff}(N,\alpha)\}; \quad H_{eff}=N-(\alpha/\delta)ln(1+\delta N/v).$$

The scaled coefficient $T$, which is an analogue for the temperature, is taken here to be unity, $T=1$. Then

$E=N$; $\beta=1$; $1/(q-1)=-\alpha/\delta$; $(q-1)r_0=\delta/v$; $r_0=-\alpha/v$; $\beta^0=1-\alpha/v$; $\beta_q=r_0=-\alpha/v$; $r_f=\beta_q=r_0=-\alpha/v$.

The relation to cases 3c), 3d) is markedly seen. In [27] the connection of the Maltus-Ferhuelst process to the finiteness of the lifetime is stressed (the population dies out with probability unity if the time is sufficiently long), and average lifetimes and survival probabilities are determined. The $\Omega$ - decomposition in [27], [28] describes the evolution on smalles time scales because the (small) probability of the anomalously large fluctuation leading to extinction $n=0$ is there neglected.

3 g) Van der Pol-Duffing system

In [24] a Van der Pol-Duffing system is considered, that is the Brownian motion in a generator with nonlinear frequency (subsection 3 c). Its effective Hamilton function is

$$H_{eff}(x,v,\alpha_f,a_f)=(M/2)[v^2-(\alpha_f/\delta)ln(1+(\delta/v)v^2)]+(M\omega_0^2/2)[x^2-(a_f/b)ln(1+bx^2)]. \qquad (27)$$

In this system the dissipative nonlinearity of the Van der Pol generator is combined with the nondissipative nonlinearity of an oscillator. The function (27) depends on two control parameters: the feedback $\alpha_f$ and field efficiency parameter $a_f$. If both controlling parameters equal to zero the effective Hamiltonian (27) then coincides with the Hamilton of the linear oscillator (linear oscillating circuit in the generator)

$$H_{eff}(x,v,\alpha_f=0, a_f=0)=(M/2)v^2+(M\omega_0^2/2)x^2=H_0.$$

The stationary solution, which is the solution of the corresponding generalized Fokker-Planck equation,

$$p_0(x,v,\alpha_f,a_f)=Cexp\{-H_{eff}/k_BT\},$$

is written as

$$p=exp\{-H_0/k_BT\}(1+(\delta/v)v^2)^{M\alpha_f/2\delta k_BT}(1+bx^2)^{M\omega_0^2 a_f/2bk_BT}. \qquad (28)$$

Since $H_0=E_1+E_2$, it is possible to derive (28) from the general expression (3) at $p=p_1p_2$;

$$p_i(x)=x^{c_i-1}exp\{-x/b_i\}/\Gamma(c_i)b_i^{c_i}; \quad i=1,2. \qquad (29)$$

We assume that the domains and variables of integration are separable, and conditions hold at which $\int_0^\infty p_1(y_1)p_2(y_1)exp\{-y_1(r_{01}E_1+r_{02}E_2)\}dy_1 = \int_0^\infty p_1(y_1)exp\{-y_1r_{01}E_1\}dy_1 \times \int_0^\infty p_2(y_1)exp\{-y_1r_{02}E_2\}dy_1$. Lets take following values of the parameters: $c_1=1/(q_1-1)$; $c_2=1/(q_2-1)$; $c_1b_1=r_{01}$; $c_2b_2=r_{02}$.

This problem can be considered as an example of not only using the gamma- and Tsallis distributions, but for taking advantages of the superstatistics, although for the mere product of gamma-distributions. The examples of this kind can be numerous.

Comparing (28) with the expression obtained from (3) at the assumption (29):

$p\sim exp\{-\beta E\}[1+(q_1-1)r_{01}E_1]^{-1/(q_1-1)}[1+(q_2-1)r_{02}E_2]^{-1/(q_2-1)}$; $H_0=E_1+E_2$; $E_1=(M/2)v^2$; $E_2=(M\omega_0^2/2)x^2$.

We get $-1/(q_1-1)=M\alpha_f/2\delta k_BT$; $-1/(q_2-1)=M\omega_0^2 a_f/2bk_BT$; $\beta=1/k_BT$; $(q_1-1)r_{01}M/2=\delta/v$; $(q_2-1)r_{02}M\omega_0^2/2=b$; $r_{01}=-\alpha_f\beta/v$; $r_{02}=-a_f\beta$.

Assuming that $\beta^0=\beta^0_1=\beta^0_2$; $r_0=r_{01}+r_{02}$, one gets: $r_0=-\beta\alpha_f(1+1/v)$; $\beta^0=(1-a_f-\alpha_f/v)/k_BT$; $\beta_{q1}=r_{01}=-\alpha_f/vk_BT$; $\beta_{q2}=r_{02}=-a_f/k_BT$; $\beta_q=\beta_{q1}+\beta_{q2}=-\beta(a_f+\alpha_f/v)$; $r_f=-\beta(a_f+\alpha_f/v)$.

At nonzero values of controlling $\alpha_f$, $a_f$ the relaxation processes in a system exhibit vast variety and complexity. In [24] two groups of different relaxation times are introduced. This opens wide possibilities of statistical description of mutual influence of nonlinear and nondissipative processes. For example, the velocity distribution at nonzero feedback essentially differs from the equilibrium one. In the stationary state (28) the number of particles with big velocities can essentially exceed the equilibrium value. This could influence strongly the transition rates of a Brownian particle in a bistable element. Thus we arrive back to the classical Brownian problem of reaching boundary time, that is to the lifetime question. The use of (28) opens new possibility of changing the transition rate – through controlling the feedback parameter in the generator. Changing the effective field the another parameter, $a_f$ changes as well. Thus the choose of some optimizing strategy is possible by changing both parameters of different nonlinearity types. Besides fluctuative transition between wells the change of the effective potential relief is also possible.

Besides the correlations (19), (20), (22), (24), (26), (28) many examples of distributions which are reduced to expressions of a kind (3) - (4) can be deduced.

## 4. Physical examples of the application of the obtained distributions

We used distribution of a kind (4) for the description self-organized criticality (*SOC*) [29]. In work [30] formation of a stationary single avalanche and fluktuated formation of an avalanche in view of additive noise for a component of speed and an inclination of sand are considered. Thus distribution of the order parameter acquires a power-like form with an integer parameter. But generally this parameter should be fractional, therefore in [30] generalization of system of Lorentz is performed allowing to account for the behaviour of ensemble of avalanches. Stochastic systems for not additive ensemble of avalanches are used also. We have performed calculation on distribution (4) with effective energy

$$E \to U(s) = \ln[(I_\Sigma + I_\zeta s^\tau)/(1+s^\tau)^2] + \int_0^s [u - u^{\tau/2}/(1+u^\tau)][(1+u^\tau)^2/(I_\Sigma + I_\zeta u^\tau)] du$$

from [30] for a case fluktuated formations of avalanches with an integer exponent of distribution of parameter of the order $\tau=2$ also have compared behaviour of stationary distribution to the results received in [30] (*Fig. 22* in [30]) for a case $\tau=1,5$ in distribution $exp\{-U(s)\}/Z$. The results are shown in *Fig.2*. Calculation was carried out for the same values of parameters noise intensivity of energy $I_\Sigma$ and complexity $I_\zeta$ which are used in [30]. Concurrence of the results, similar behaviour of distributions is evidenced. Parameters $q$, $r_0$ only weakly influence behaviour of distribution. We shall note, that using the pure Tsallis distribution leads to another behaviour of stationary distribution (in particular, does not show indicated in [30] distinctions between distributions with $I_\Sigma=1$; $I_\zeta=5$ and $I_\Sigma=0,5$; $I_\zeta=30$).

Comparison with the results obtained [31, 32] for a spectrum of the cosmic ray was carried out as well. By means of distribution of a kind $p(E)=CE^2(1+(q-1)\tilde{\beta} E)^{-1/(q-1)}$ (*C* is a constant representing the total flux rate; the Tsallis distribution is multiplied with $E^2$, taking into account the available phase space volume) at $q=1,215$, $\tilde{\beta}^{-1}=107$ *MeV*, $C=5 \times 10^{-13}$ in [31] the measured flux rate of cosmic ray particles with a given energy is well fitted. If we use for these purposes the distribution (4) multiplied with $CE^2$ there is an additional parameter $r_0=p_0/u$ (*Fig.3*). At values $p_0/u$ close to *1* we get coincidence to results of [31]. At $p_0/u=1-10^{-14}$ the coincidence to results [31] is exact. At $p_0/u=1-10^{-13}$ sharper bend is obtained at ankle energy $\sim 10^{19}$ *eV*; the plot for flux rate comes nearer to curve for parameters $q=11/9$, $C=10^{-14}$ from [31] as well. Decreasing $p_0/u$ we get the dependence, characteristic for Boltzmann-Gibbs distribution, resulted on *Fig. 1* in [32]. At values $p_0/u \approx 1$ it is possible to explain conformity of experimental data in small volume of system both high energy and temperatures. Comparison of calculations was carried out also by means of expression (4) with estimations of distributions for the motion of point defects in thermal convection patterns in an inclined fluid layer [4], [33] which also has shown conformity with the results received in [4], [33].

The same distributions (but in a discrete variant as in [1]) describe distribution on energy of density of neutrons in a nuclear reactor [1], and in a continuous variant - distribution of charges in the charged aerosols. In both these phenomena essential is the finiteness of a lifetime of elements (neutrons and particles of an aerosol) in comparison with an environment (nucleus of moderator for a reactor and molecules of air for aerosol particles).

## 3. Conclusion

In our work we have obtained the superstatistics in a way different from that of [5]. This way is free from the shortcomings pointed out in [34]. Deriving (5) of [1] the notion of the slave process [35] was used which plays part for the derivation of superstatistics as well [36]. The distributions (1-4) are products of the Boltzmann factor and the superstatistics factor. The Boltzmann factor causes them to behave as Gibbs distributions (*Fig. 1*) at $0<r_0<1$. However at $r_0<0$ and $r_0>1$ the distributions behave very differently. Thus these distributions present a possible tool for phenomenological description of the bifurcations and phase transitions. The relations (14)-(29) demonstrate this using the examples of known systems. Hereto the anomaly diffusion example [37] and a number of other phenomena could be added which are characterized by similar distributions.

The second factor of the distribution (4) contains an additional parameter $p_0/u$ in comparison to the Tsallis distribution; from the examples of Section 3 one can judge about the difference of the suggested distribution from Tsallis one. So, for zero values of controlling parameter at which the distribution should pass to Gibbs one, the value *q-1* has peculiarity. The additional parameter $p_0/u$ allows for comparison of the distributions obtained with the results (17)-(28) and determining explicit values of parameters. The general form (2) provides even more possibilities. Therefrom the distributions (6)-(13) are obtained, and their applicability to various physical situations is to be elucidated. The second factor in (3) after changing $r_0 \to \beta^0$ passes to the expression for superstatistics [5]. The suggested approach can be considered as broadening and detalizing the superstatistics theory. Three characteristic temperatures are named: Gibbs temperature (8) of [1] $\beta=\beta_{Gibbs}=1/k_B T=\beta^0(1-p_0/u)$; effective temperature $\beta^0 =\beta_{eff}=\beta/(1-p_0/u)$; and Tsallis temperature $\beta_q=\beta_{Ts}=r_0=\beta(p_0/u)/(1-p_0/u)$; the relations are derived between them and the parameters $r_0$, $r_f$: $\beta^0=\beta+\beta_q$; $r_f= r_0=<\alpha P/u>$.

In [38] one more interesting question on a form of the differential equations for $y(x)=p(E)$ is considered. For distribution (4) this equation looks like $dp(E)/dE=-(1-r_0)p(E)-r_0 p(E)/[1+(q-1)r_0 E]$. It is possible to lead analogies to the equations, which have been written down in [38]. It is possible for this reason gamma-distributions $f(x)=\lambda^\alpha x^{\alpha-1} exp\{-\lambda x\}/\Gamma(\alpha)$ with a parameter $\alpha=r/2$ (is more true, $\chi^2$ distributions with *r* degrees of freedom which corresponds to gamma-distribution at $\lambda=1/2$-c; in [5] value *r* is interpreted as number of degrees of freedom giving the contribution in fluktuating value $\beta$) precisely describe Tsallis statistics, - they correspond to distribution of lifetime with *r* stages.

The Gibbs distribution does not describe the dissipative processes that develop in the system. Superstatistics describe systems by constantly putting energy into the system, which is dissipated. Similarly in [39] the new interpretation of the entropic index *q* has been found. In [39] it has been shown that in $q=(c+2)/(c+1)$ (as function of the negative-binomial distribution) parameter *c* is related to the number of random, exponentially distributed contributions to the effective statistics of a complex system. But if the value $q$ determines Gibbs contribution to the superstatistics, the parameter $r_0$ is in its turn responsible for the relation between two factors in (3).

It is possible not to pass to integrated relations, as in (3), having limited to summation (so, as discrete analogue of gamma-distribution negative binomial distribution serves [1]). The discrete description in many cases, for example, for bistability potential, appears more precisely continuous. The found conformity between superstatistics and the nonequilibrium distribution containing lifetime, should appear useful to both cases. For example, many results established for superstatistics and nonextensive statistical mechanics, are transferred to the description of complex systems by means of the distribution containing lifetime. Since thermodynamics with lifetime [40,41,42] is more general, than the theory of superstatistics also it has more opportunities. Interesting is establishing the relation to a method of the nonequilibrium statistical operator of Zubarev [2] generalizing Gibbs distributions which as it was marked, it is possible to compare to the distributions containing lifetime, and nonextensive statistical mechanics [6-7], in which entropy is represented by means of the measures which are distinct from Boltzmann and Gibbs. Probably, they stand close to the concept of lifetime, in the latent kind present in both theories. The obtained distribution contains the new parameter related to a thermodynamic state of the system, and also with distribution of a lifetime of a metastable states and interaction of this states with an environment. Changing this parameter it is possible to pass to Gibbs distribution.

Thus, an explicit account for the lifetime finiteness for a statistical system and the use of the lifetime (first-passage time) as a thermodynamic variable brings us to the nonextensive statistical mechanics. This fact is in a correspondence with known results about the finiteness of the size for systems in which the nonextensive statistical mechanics can be applied [43]. Furthermore we obtained the distribution generalizing the Gibbs-Boltzmann statistics and superstatistics. Real open systems seem to share the properties of both Gibbs and nonextensive statistical mechanics types. So we believe bringing forth such kinds of distributions to be a perspective point of view, which is likewise confirmed by the relevance of such distributions for various phenomena. Depending on the value of $r_0$ and on the controlling parameter the properties of either nonextensive or extensive statistical mechanics prevail. The obtained relation (3) can be regarded as the detalization (concretization) of the superstatistics theory [5]. For the expression (3) one can write the entropy expression and corresponding thermodynamic relations as well.

# Figure captions

Fig.1. Comparison of distributions (4) $f(v)=\exp\{-\beta v^2/2\}Z^{-1}(1+(q-1)r_0v^2/2)^{-1/(q-1)}$ ($E=v^2/2$; $r_0=0,1$; $\beta=0,1$; $q=1,1$) and $p(v)=\exp\{-\beta v^2/2\}[1+(q-1)(1-\exp\{-\beta(p_0/u)v^2/2(1-p_0/u)\}]^{-1/(q-1)}/Z_1$ with Boltzmann-Gibbs distribution $n(v)=\exp\{-\beta v^2/2\}/Z_B$ and with Tsallis distribution $t(v)=[1+(q-1)r_0v^2/2]^{1/(1-q)}/Z_T$.

Fig.2. Function of distribution (4) in the description self-organized criticality (SOC) at $\tau=2$, $q-1=4/11$, $r_0=0,1$; $P(s)$ $(R(s), f(s))=[1+0,1(q-1)U(s)]^{-1/c}\exp\{-0,9U(s)\}$; $U(s)=\ln[(I_\Sigma+I_\zeta s^\tau)/(1+s^\tau)^2]+\int_0^s [u-u^{\tau/2}/(1+u^\tau)][(1+u^\tau)^2/(I_\Sigma+I_\zeta u^\tau)]du$; $P(s)$: $I_\Sigma=0$; $I_\zeta=50$; $R(s)$: $I_\Sigma=0,5$; $I_\zeta=30$; $f(s)$: $I_\Sigma=1$; $I_\zeta=5$.

Fig.3. Energy spectrum of primary cosmic rays (in units of $m^{-3}s^{-1}sr^{-1}GeV^{-1}$) as listed in [31]. The line $p(E)$ is the prediction by $p(E)=CE^2(1+(q-1)\tilde{\beta}E)^{-1/(q-1)}$ with $q=1,215$, $\tilde{\beta}^{-1}=k_B\tilde{T}=107\ MeV$, $C=5\times10^{-13}$ in the above units [31]; values of energy are specified in $eV$. The function $f(E)=CE^2\exp\{-\tilde{\beta}(1-r_0)E\}(1+(q-1)\tilde{\beta}(p_0/u)E)^{-1/(q-1)}$ obtained on (4) with $p_0/u=1-10^{-14}$. Functions $g(E)$, $k(E)$, $r(E)$, $s(E)$, $e(E)$ too are obtained on expression (4) with $p_0/u=1-10^{-13}$, $p_0/u=1-10^{-12}$, $p_0/u=1-10^{-11}$, $p_0/u=1-10^{-10}$, $p_0/u=1-10^{-9}$ accordingly.

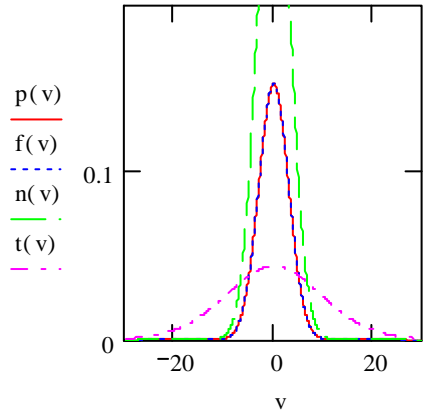

Fig.1.

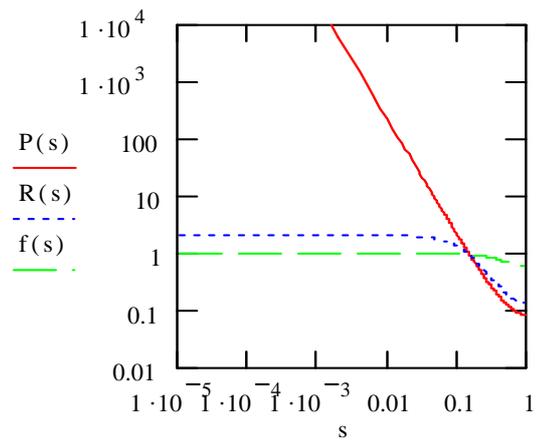

Fig.2.

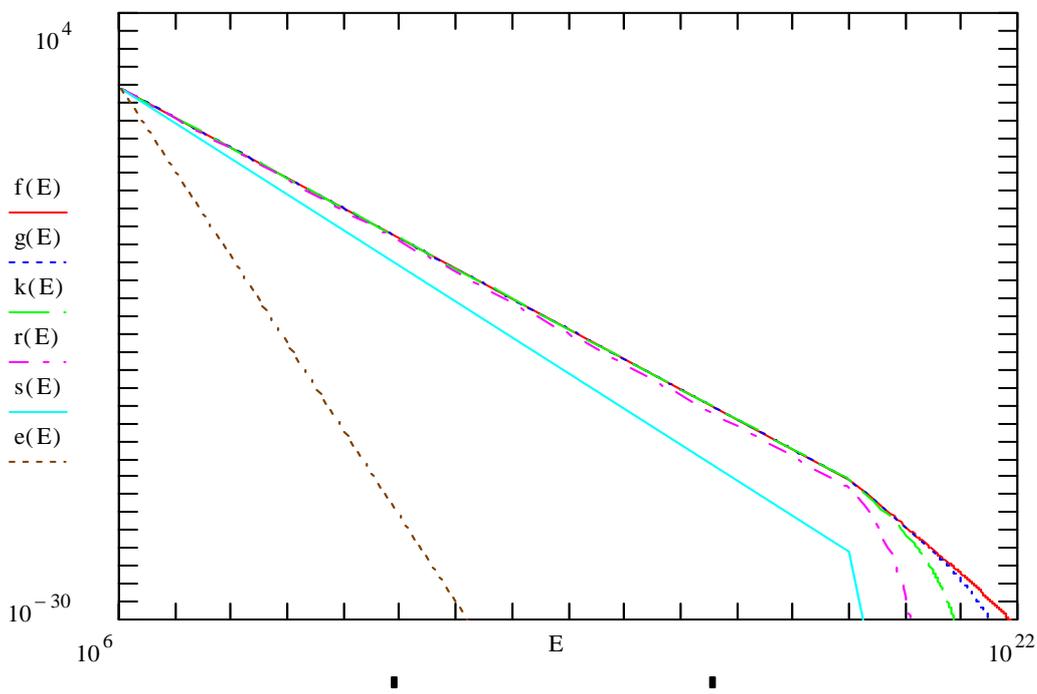

Fig.3.